\documentclass[Physsubmission, Phys]{SciPost}

\hypersetup{
    colorlinks,
    linkcolor={red!50!black},
    citecolor={blue!50!black},
    urlcolor={blue!80!black}
}

\usepackage[bitstream-charter]{mathdesign}
\newcommand{\eloise}{\textsc{Eloise}}
\newcommand{\nucleus}{\textsc{Nucleus}}
\newcommand{\cresst}{\textsc{Cresst}}
\newcommand{\crab}{\textsc{Crab}}
\usepackage{textgreek}
\newcommand{\cevns}{CE\textnu{}NS}

\usepackage{cleveref}
\usepackage{siunitx}
\usepackage[version=4]{mhchem}
\newcommand{\cawo}{\ce{CaWO_4}}
\newcommand{\alo}{\ce{Al_2O_3}}
\usepackage{booktabs}
\usepackage{subfig}

\DeclareSymbolFont{usualmathcal}{OMS}{cmsy}{m}{n}
\DeclareSymbolFontAlphabet{\mathcal}{usualmathcal}

\begin{document}

\begin{center}{\Large \textbf{
ELOISE – Reliable background simulation at sub-keV energies\\
}}\end{center}

\begin{center}
Holger Kluck\textsuperscript{1$\star$}
\end{center}

\begin{center}
{\bf 1} Institut für Hochenergiephysik der Österreichischen Akademie der Wissenschaften, A-1050 Wien, Austria
\\
* holger.kluck@oeaw.ac.at
\end{center}

\begin{center}
\today
\end{center}


\definecolor{palegray}{gray}{0.95}
\begin{center}
\colorbox{palegray}{
  \begin{tabular}{rr}
  \begin{minipage}{0.1\textwidth}
    \includegraphics[width=30mm]{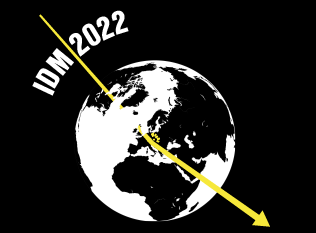}
  \end{minipage}
  &
  \begin{minipage}{0.85\textwidth}
    \begin{center}
    {\it 14th International Conference on Identification of Dark Matter}\\
    {\it Vienna, Austria, 18-22 July 2022} \\
    \doi{10.21468/SciPostPhysProc.?}\\
    \end{center}
  \end{minipage}
\end{tabular}
}
\end{center}

\section*{Abstract}
{\bf

\cawo{} and \alo{} are well-established target materials used by experiments searching for rare events like the elastic scattering off of a hypothetical Dark Matter particle. In recent years, experiments have reached detection thresholds for nuclear recoils at the \SI{10}{\eV}-scale. At this energy scale, a reliable Monte Carlo simulation of the expected background is crucial. However, none of the publicly available general-purpose simulation packages are validated at this energy scale and for these targets. The recently started \eloise{} project aims to provide reliable simulations of electromagnetic particle interactions for this use case by obtaining experimental reference data, validating the simulation code against them, and, if needed, calibrating the code to the reference data.
}

\vspace{10pt}
\noindent\rule{\textwidth}{1pt}
\tableofcontents\thispagestyle{fancy}
\noindent\rule{\textwidth}{1pt}
\vspace{10pt}

\section{Introduction}
\label{sec:intro}
The \eloise{} (R\textbf{el}iable Backgr\textbf{o}und S\textbf{i}mulation at \textbf{S}ub-keV \textbf{E}nergies) project aims to validate and, if necessary, to tune and extend the Monte Carlo (MC) simulation of electromagnetic interactions in \cawo{} and \alo{} at sub-keV energies of the free particle approximation. I first motivate the necessity for a reliable simulation in these materials and at this energy scale in \cref{sec:motivation}. Afterwards, relevant physics processes for \eloise{} in \cref{sec:scope} are identified and suitable MC codes are reviewed in \cref{sec:mccodes}. In \cref{sec:method}, the applied methodology is outlined. The current status of \eloise{} is reported in \cref{sec:status} before I conclude in \cref{sec:conclusion}.

\section{Importance of a reliable background simulation}
\label{sec:motivation}

\cawo{} and \alo{} are widely used target materials for cryogenic calorimeters \cite{Angloher2019,Abdelhameed2019,Thulliez2021}. Operated at mK-temperatures, the measured phonon signal allows for detection of nuclear recoils which are the signature for highly interesting \textit{rare} particle processes:
within the Standard Model (SM) of particle physics, the Coherent Elastic Neutrino-Nucleus Scattering (\cevns{}) is known to cause nuclear recoils. The \nucleus{} experiment \cite{Angloher2019} aims to measure \cevns{} mainly with \cawo{} calorimeters as a new probe for electro-weak precision measurements. New Physics beyond the SM may provide a particle candidate for Dark Matter (DM), one of the greatest mysteries of modern physics. The \cresst{} experiment \cite{Abdelhameed2019} searches for nuclear recoils  caused by the elastic scattering of hypothetical DM particles off the nuclei in \cawo{} and \alo{} among other target materials. Beyond basic research, the \crab{} project \cite{Thulliez2021} aims to develop a new energy calibration technique for nuclear recoils in \cawo{} based on thermal neutron capturing.

In all three cases, the respective energy scales go down to the sub-keV regime: \nucleus{} and \cresst{} demonstrated detection threshold for nuclear recoils of \SI{19.7}{\eV} \cite{Strauss2017} and \SI{30.1}{\eV} \cite{Abdelhameed2019}, respectively. \crab{} aims for a calibration signal as low as \SI{112.5}{\eV} \cite{Thulliez2021}.

At these energies, nearly all\footnote{For scintillating \cawo{} targets, it is generally possible to separate nuclear recoils from other types of interactions via the scintillation light yield. However, at the \SI{10}{\eV}-scale the effectiveness of this approach degenerates quickly \cite{Abdelhameed2019,Abdelhameed2019b}.} particle interactions that deposits energy in the respective energy range of interest cause background events. The identification of the sought-after rare events against this dominant background relies crucially on a reliable background model.

Further interest in the detailed physics at lowest energies was sparked by the so-called \textit{low-energy excess}: the observation by \cresst{}, \nucleus{}, and others of a yet unexplained increase of events with decreasing energies in the sub-keV regime, constituting an excess above the established background \cite{Adari2022}.

\section{Relevant physics processes}
\label{sec:scope}
\begin{figure}[h]
\centering
\subfloat[][]{\includegraphics[width=0.4\textwidth]{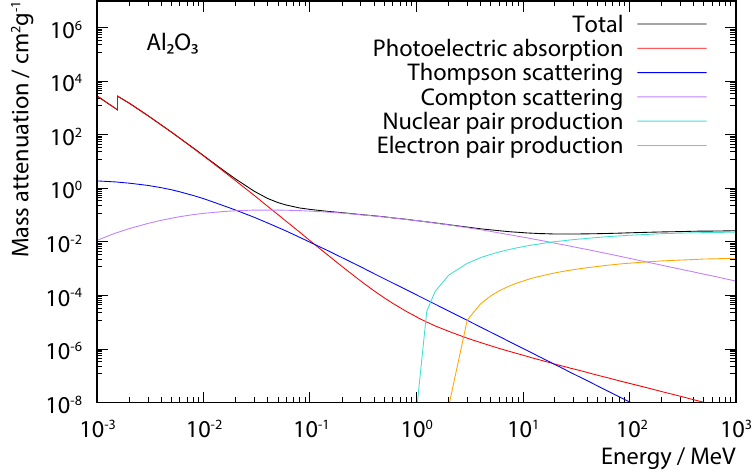}}
\qquad
\subfloat[][]{\includegraphics[width=0.4\textwidth]{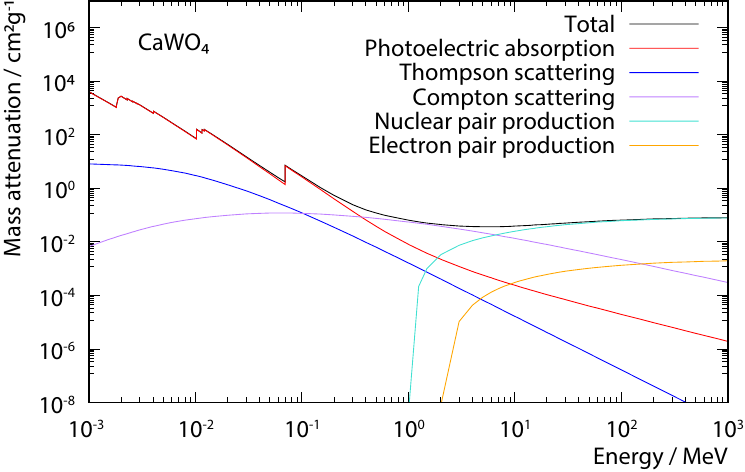}}\\
\subfloat[][]{\includegraphics[width=0.4\textwidth]{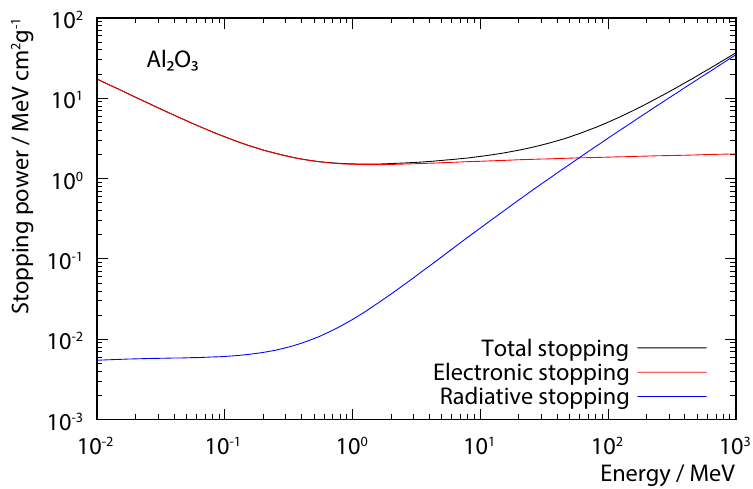}}
\qquad
\subfloat[][]{\includegraphics[width=0.4\textwidth]{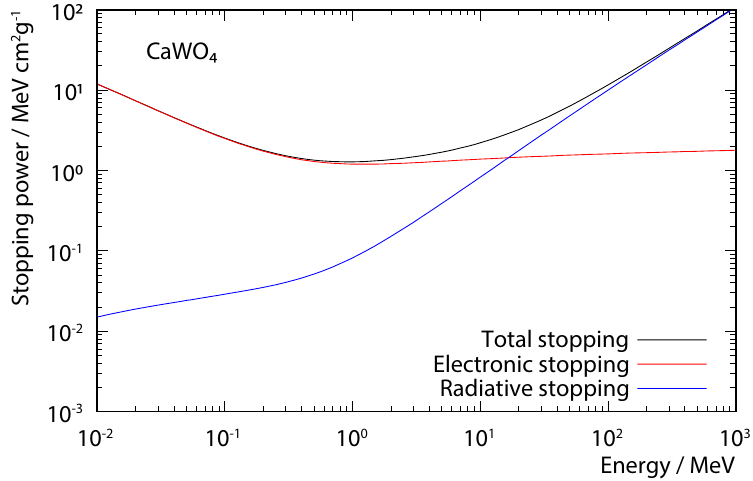}\label{fig:estar}}
\caption{\label{fig:energyloss}Literature data for attenuation of photons \cite{xcom} (\textit{top} row) and stopping power of electrons \cite{estar} (\textit{bottom} row) in \alo{} (\textit{left} column) and \cawo{} (\textit{right} column).}
\end{figure}

\Cref{tab:quantities} lists the maximum recoil energies caused by \cevns{} or elastic DM-nucleus scattering for two typical benchmark cases. 
\begin{table}
\caption{\label{tab:quantities}Calculated maximum recoil energies caused by \cevns{} with a neutrino of \SI{2}{\MeV} kinetic energy ($E_\mathrm{rec,\nu}$) and by elastic scattering with a $2\,\mathrm{GeV/c^2}$-DM particle with a velocity of \SI{220}{\meter\per\second} ($E_\mathrm{rec,DM}$) and the minimum displacement energies ($E_\mathrm{dis}$) for the nuclides of \cawo{} \cite{Shao2008} and Al of \alo{} \cite{Yuan2017}.}
\centering
\begin{tabular}{l S S S S}\toprule
                              & {\ce{_8O}} & {\ce{_13Al}} & {\ce{_20Ca}} & {\ce{_74W}} \\ \midrule
$E_\mathrm{rec,DM}/\si{\eV}$  & 106.4      &  69.2        &   48.6       & 11.5  \\
$E_\mathrm{rec,\nu}/\si{\eV}$ & 499.9      & 296.5        &  199.5       & 43.5  \\ 
$E_\mathrm{dis}/\si{\eV}$     & 20         &  47.5        &   24         & 196 \\\bottomrule
\end{tabular}
\end{table}
If the recoil energy ($E_\mathrm{rec}$) is below the displacement energy ($E_\mathrm{dis}$), as for W in the benchmark cases, it is transformed to electron and phonon excitations of the crystal lattice. In case of $E_\mathrm{rec} > E_\mathrm{dis}$, as is the case for O, Al, and Ca, the atom leaves its lattice site as Primary Knock-on Atom (PKA) and creates a vacancy (see e.g. \cite{Kinchin1955}). In general, $E_\mathrm{dis}$ depends on the orientation of the PKA relative to the crystal lattice structure \cite{Yuan2017}. Via nuclear and electronic stopping, the PKA loses kinetic energy until it stops as an interstitial atom and causes a crystal defect. At the expected PKA energies, the displaced nucleus loses its kinetic energy mainly via \textit{Coulomb scattering} (nuclear stopping) \cite{PDG2018}. If enough energy is transferred to secondary atoms, a nuclear recoil cascade may form.  As the target crystals are operated at mK-temperatures, a recovery of the caused damage is strongly suppressed.
The PKA and secondary atoms may de-excite via \textit{fluorescence} and emission of \textit{Auger electrons}. At a tertiary stage, the X-rays and electrons may lose their energies mainly via \textit{photoelectric absorption} and \textit{ionisation}, respectively (cf.\ \cref{fig:energyloss}).

Since the neutron-induced background is also ultimately converted into electromagnetic interactions via the neutron-nucleus scattering and the resulting recoiling nucleus, \eloise{} will be focused on the highlighted electromagnetic processes.

\section{Suitable Monte Carlo packages}
\label{sec:mccodes}

Due to the wide range of involved particles and processes as well as the complex detector geometry, the background of experiments searching for rare events is usually modelled with a general purpose MC package like Geant4 \cite{Allison2016}, FLUKA \cite{Boehlen2014}, or MCNP \cite{Werner2018}.
Albeit these codes were validated for various materials and energies, none of them has been validated specifically for \cawo{} and \alo{} at sub-keV energies. As pointed out in \cite{Basaglia2015a, Basaglia2016a}, it is usually precarious to extrapolate the correctness for a given use case from studies based on a different use case. As all three codes are based on the \textit{free particle approximation} of cold, neutral, and unbound atoms, only a dedicated validation can determine if this approximation is still suitable at the \SI{10}{\eV}-scale in \cawo{} and \alo{}. Based on the displacement energies listed in \cref{tab:quantities}, I expect the onset of significant solid state-effects and hence the break-down of the free particle approximation at roughly \SI{50}{\eV}.

Out of the listed MC codes, the recommended application limit\footnote{See \url{https://geant4.web.cern.ch/node/1619}.} of Geant4 is the lowest with \SI{250}{\eV}. Furthermore, its source code is publicly available and well documented. This has already enabled the user community to extend Geant4’s applicability down to the \SI{10}{\eV}-scale in crystalline Si (“MicroElec” project \cite{Raine2014}). Hence, I adopted Geant4 for my studies.

\section{Validation protocol}
\label{sec:method}

To provide reliable simulations of the electromagnetic processes in \cawo{} and \alo{} from \SI{1}{\keV} down to the onset of solid state-effects, I will apply a four-step approach: (i) collect experimental reference data, (ii) run Geant4 based MC simulations of the reference measurements, (iii) validate the simulation against the references data, and if needed (iv) tune the existing MC models to the reference data or, if this is not sufficient, develop new models. 

Step (i) will be realised either via literature searches or via dedicated measurements conducted within \eloise{}. To minimise systematic errors, step (ii) requires a detailed simulation of the experimental measurement. In step (iii) the methodology developed by T.\ Basaglia et al.\ (see e.g. \cite{Basaglia2016a}) will be applied for the validation of Geant4 based MC simulations to provide an objective assessment. If the development of new, dedicated physics models is needed in step (iv), it is planned to adapt the approach of the MicroElec project \cite{Raine2014}.

\section{Status of \eloise{}}
\label{sec:status}
At the moment, I am studying the energy loss of electrons in \cawo{} by ionisation. As indicated in \cref{fig:estar}, the commonly tabulated reference data goes down only to \SI{10}{\keV}. Hence, in step (i) a dedicated electron energy loss spectroscopy (EELS, see e.g.\ \cite{eels}) of a \SI{105}{\nm} \textit{thin} \cawo{} sample\footnote{To reduce complication by multiple electron interactions in a \textit{thick} sample, a \textit{thin} sample was prepared to collect data of mostly single electron interactions as reference. The precise sample thickness was not determined a priori but is an outcome of the sample preparation. However, it has to be precisely considered in the Geant4 simulation of the measurement to ensure compatibility.} was ordered. The obtained data in the energy range from \SIrange{0}{2}{\keV} are presently prepared for publication. In step (ii) I am currently running a detailed simulation of the measurement based on Geant4 version 10.6.3.

\section{Conclusion}
\label{sec:conclusion}
Experiments searching at the \SI{10}{\eV}-scale for nuclear recoils as signature for rare events depend crucially on a reliable MC simulation of their background. At this energy scale, only electromagnetic interactions contribute to the known background. As the breakdown of the free particle approximation in \cawo{} and \alo{} can be expected at $\approx \SI{50}{\eV}$, a validation of the MC codes is necessary. Currently, \eloise{} prepares the validation of Geant4 for  electron ionisation in \cawo{} against a dedicated reference measurement in the energy range of \SIrange{0}{2}{\keV}.

\paragraph{Funding information}
The \eloise{} project and HK are funded by the Austrian Science Fund (FWF): P 34778-N "ELOISE".

\bibliography{IDM2022_HolgerKluck_ELOISE.bib}

\begin{thebibliography}{10}
\providecommand{\url}[1]{\texttt{#1}}
\providecommand{\urlprefix}{URL }
\expandafter\ifx\csname urlstyle\endcsname\relax
  \providecommand{\doi}[1]{doi:\discretionary{}{}{}#1}\else
  \providecommand{\doi}{doi:\discretionary{}{}{}\begingroup
  \urlstyle{rm}\Url}\fi
\providecommand{\eprint}[2][]{\url{#2}}

\bibitem{Angloher2019}
G.~Angloher, F.~Ardellier-Desages, A.~Bento, L.~Canonica, A.~Erhart,
  N.~Ferreiro, M.~Friedl, V.~M. Ghete, D.~Hauff, H.~Kluck, A.~Langenk{\"a}mper,
  T.~Lasserre \emph{et~al.},
\newblock \emph{Exploring {CE$\nu$NS} with {NUCLEUS} at the {Chooz} nuclear
  power plant},
\newblock Eur. Phys. J. C \textbf{79}(12), 1018 (2019),
\newblock \doi{10.1140/epjc/s10052-019-7454-4}.

\bibitem{Abdelhameed2019}
A.~H. Abdelhameed, G.~Angloher, P.~Bauer, A.~Bento, E.~Bertoldo, C.~Bucci,
  L.~Canonica, A.~D'Addabbo, X.~Defay, S.~Di~Lorenzo, A.~Erb, F.~v. Feilitzsch
  \emph{et~al.},
\newblock \emph{First results from the {CRESST-III} low-mass dark matter
  program},
\newblock Phys. Rev. D \textbf{100}, 102002 (2019),
\newblock \doi{10.1103/PhysRevD.100.102002}.

\bibitem{Thulliez2021}
L.~Thulliez, D.~Lhuillier, F.~Cappella, N.~Casali, R.~Cerulli, A.~Chalil,
  A.~Chebboubi, E.~Dumonteil, A.~Erhart, A.~Giuliani, F.~Gunsing, E.~Jericha
  \emph{et~al.},
\newblock \emph{Calibration of nuclear recoils at the 100 {eV} scale using
  neutron capture},
\newblock J. Instrum. \textbf{16}(07), P07032 (2021),
\newblock \doi{10.1088/1748-0221/16/07/p07032}.

\bibitem{Strauss2017}
R.~Strauss \emph{et~al.},
\newblock \emph{{The $\nu$-cleus experiment: A gram-scale fiducial-volume
  cryogenic detector for the first detection of coherent neutrino-nucleus
  scattering}},
\newblock Eur. Phys. J. C \textbf{77}, 506 (2017),
\newblock \doi{10.1140/epjc/s10052-017-5068-2}.

\bibitem{Abdelhameed2019b}
A.~H. Abdelhameed, G.~Angloher, P.~Bauer, A.~Bento, E.~Bertoldo, R.~Breier,
  C.~Bucci, L.~Canonica, A.~D'Addabbo, S.~D. Lorenzo, A.~Erb, F.~v. Feilitzsch
  \emph{et~al.},
\newblock \emph{Geant4-based electromagnetic background model for the {CRESST}
  dark matter experiment},
\newblock Eur. Phys. J. C \textbf{79}(10), 881 (2019),
\newblock \doi{10.1140/epjc/s10052-019-7385-0},
\newblock [Erratum: Eur. Phys. J. C \textbf{79}(12), 987 (2019),
  doi:\href{https://doi.org/10.1140/epjc/s10052-019-7504-y}{10.1140/epjc/s10052-019-7504-y}].

\bibitem{Adari2022}
P.~Adari, A.~Aguilar-Arevalo, D.~Amidei, G.~Angloher, E.~Armengaud, C.~Augier,
  L.~Balogh, S.~Banik, D.~Baxter, C.~Beaufort, G.~Beaulieu, V.~Belov
  \emph{et~al.},
\newblock \emph{{EXCESS} workshop: Descriptions of rising low-energy spectra},
\newblock SciPost Phys. Proc. p. 001 (2022),
\newblock \doi{10.21468/SciPostPhysProc.9.001}.

\bibitem{Shao2008}
Z.~Shao, Q.~Zhang, T.~Liu and J.~Chen,
\newblock \emph{Computer study of intrinsic defects in {$CaWO_4$}},
\newblock Nucl. Instrum. Methods Phys. Res. B \textbf{266}(5), 797 (2008),
\newblock \doi{10.1016/j.nimb.2008.01.018}.

\bibitem{Yuan2017}
Y.~G. Yuan, M.~Jiang, F.~A. Zhao, H.~Chen, H.~Gao, H.~Y. Xiao, X.~Xiang and
  X.~T. Zu,
\newblock \emph{Ab initio molecular dynamics simulation of low energy radiation
  responses of $\alpha$-{$Al_2O_3$}},
\newblock Sci. Rep. \textbf{7}(1), 3621 (2017),
\newblock \doi{10.1038/s41598-017-03827-1}.

\bibitem{Kinchin1955}
G.~H. Kinchin and R.~S. Pease,
\newblock \emph{The displacement of atoms in solids by radiation},
\newblock Rep. Prog. Phys. \textbf{18}(1), 1 (1955),
\newblock \doi{10.1088/0034-4885/18/1/301}.

\bibitem{PDG2018}
M.~Tanabashi, K.~Hagiwara, K.~Hikasa, K.~Nakamura, Y.~Sumino, F.~Takahashi,
  J.~Tanaka, K.~Agashe, G.~Aielli, C.~Amsler, M.~Antonelli, D.~M. Asner
  \emph{et~al.},
\newblock \emph{Review of particle physics},
\newblock Phys. Rev. D \textbf{98}, 030001 (2018),
\newblock \doi{10.1103/PhysRevD.98.030001}.

\bibitem{xcom}
M.~Berger, J.~Hubbell, S.~Seltzer, J.~Chang, J.~Coursey, R.~Sukumar, D.~Zucker
  and K.~Olsen,
\newblock \emph{{XCOM}: Photon cross sections database},
\newblock Tech. Rep. {NIST} Standard Reference Database 8 ({XGAM}), National
  Institute of Standards and Technology,
\newblock \doi{10.18434/T48G6X} (2010).

\bibitem{estar}
M.~Berger, J.~Coursey, M.~Zucker and J.~Chang,
\newblock \emph{Stopping-power \& range tables for electrons, protons, and
  helium ions},
\newblock Tech. Rep. {NIST} Standard Reference Database 124, National Institute
  of Standards and Technology,
\newblock \doi{10.18434/T4NC7P} (2017).

\bibitem{Allison2016}
J.~Allison, K.~Amako, J.~Apostolakis, P.~Arce, M.~Asai, T.~Aso, E.~Bagli,
  A.~Bagulya, S.~Banerjee, G.~Barrand, B.~Beck, A.~Bogdanov \emph{et~al.},
\newblock \emph{Recent developments in {Geant4}},
\newblock Nucl. Instrum. Methods Phys. Res. A \textbf{835}, 186 (2016),
\newblock \doi{10.1016/j.nima.2016.06.125}.

\bibitem{Boehlen2014}
T.~Böhlen, F.~Cerutti, M.~Chin, A.~Fassò, A.~Ferrari, P.~Ortega, A.~Mairani,
  P.~Sala, G.~Smirnov and V.~Vlachoudis,
\newblock \emph{The {FLUKA} code: Developments and challenges for high energy
  and medical applications},
\newblock Nucl. Data Sheets \textbf{120}, 211 (2014),
\newblock \doi{10.1016/j.nds.2014.07.049}.

\bibitem{Werner2018}
C.~J. Werner, J.~S. Bull, C.~J. Solomon, F.~B. Brown, G.~W. McKinney, M.~E.
  Rising, D.~A. Dixon, R.~L. Martz, H.~G. Hughes, L.~J. Cox, A.~J. Zukaitis,
  J.~C. Armstrong \emph{et~al.},
\newblock \emph{{MCNP} version 6.2 release notes},
\newblock Tech. Rep. LA-UR-18-20808, Los Alamos National Laboratory,
\newblock \doi{10.2172/1419730} (2018).

\bibitem{Basaglia2015a}
T.~Basaglia, M.~C. Han, G.~Hoff, C.~H. Kim, S.~H. Kim, M.~G. Pia and
  P.~Saracco,
\newblock \emph{Experimental quantification of {Geant4} {PhysicsList}
  recommendations: methods and results},
\newblock J. Phys.: Conf. Ser. \textbf{664}(7), 072037 (2015),
\newblock \doi{10.1088/1742-6596/664/7/072037}.

\bibitem{Basaglia2016a}
T.~Basaglia, M.~C. Han, G.~Hoff, C.~H. Kim, S.~H. Kim, M.~G. Pia and
  P.~Saracco,
\newblock \emph{Quantitative test of the evolution of {Geant4} electron
  backscattering simulation},
\newblock {IEEE} Trans. Nucl. Sci. \textbf{63}(6), 2849 (2016),
\newblock \doi{10.1109/tns.2016.2617834}.

\bibitem{Raine2014}
M.~Raine, M.~Gaillardin and P.~Paillet,
\newblock \emph{{Geant4} physics processes for silicon microdosimetry
  simulation: Improvements and extension of the energy-range validity up to 10
  {GeV}/nucleon},
\newblock Nucl. Instrum. Methods Phys. Res. B \textbf{325}, 97 (2014),
\newblock \doi{10.1016/j.nimb.2014.01.014}.

\bibitem{eels}
R.~Egerton,
\newblock \emph{Electron Energy-Loss Spectroscopy in the Electron Microscope},
\newblock Springer,
\newblock \doi{10.1007/978-1-4419-9583-4} (2011).

\end{thebibliography}

\nolinenumbers

\end{document}